\documentclass[manuscript]{aastex}
\begin{document}
\title{Composite Accretion Disk and White Dwarf Photosphere Analyses
of the FUSE and HST Observations of EY Cygni}

\author{Edward M. Sion, Lisa Winter, Joel A. Urban}
\affil{Dept. of Astronomy \& Astrophysics,
Villanova University,
Villanova, PA 19085}
 \email{\{edward.sion, lisa.winter, joel.urban\}@villanova.edu}
 \author{Gaghik H. Tovmassian,
         Sergei Zharikov}
\affil{Observatorio Astron\'omico Nacional, Instituto de
              Astronom\'\i a, UNAM, \\ P.O. Box 439027, San Diego,
              CA, 92143-9027 }
              \email{\{gag, zhar\}@astrosen.unam.mx}
 \author{Boris T. G\"{a}nsicke}
\affil{University of Warwick, United Kingdom}
\email{boris.gaensicke@warwick.ac.uk}
\author{M. Orio}
\affil{INAF, Osservatorio Astronomico di Torino, Strada Osservatorio
20,\\
           Pino Torinese (TO), I-10025 Italy \\ 
\and
 Astronomy Department, University of  Wisconsin, 
               475 N. Charter Str.,  Madison WI 53706} 
            \email{orio@cow.physics.wisc.edu}

\begin{abstract}

We explore the origin of FUSE and HST STIS far UV spectra of the dwarf
nova, EY Cyg, during its quiescence using \emph{combined} high gravity
photosphere and accretion disk models as well as model accretion belts.  
The best-fitting single temperature white dwarf model to the FUSE plus HST
STIS spectrum of EY Cygni has T$_{eff} = 24,000$K, log $g = 9.0$, with an
Si abundance of 0.1 x solar and C abundance of 0.2 x solar but the
distance is only 301 pc. The best-fitting composite model consists of
white dwarf with T$_{eff} = 22,000$K, log $g = 9$, plus an accretion belt
with T$_{belt} = 36,000$K covering 27\% of the white dwarf surface with
V$_{belt} sin i = 2000$ km/s. The accretion belt contributes 63\% of the
FUV light and the cooler white dwarf latitudes contribute 37\%. This fit
yields a distance of 351 pc which is within 100 pc of our adopted distance
of 450 pc. EY Cyg has very weak C\,{\sc iv} emission and very strong
N\,{\sc v} emission, which is atypical of the majority of dwarf novae in
quiescence. We also conducted a morphological study of the surroundings of
EY Cyg using direct imaging in narrow nebular filters from ground-based
telescopes. We report the possible detection of nebular material
associated with EY Cygni. Possible origins of the apparently large N\,{\sc
v}/C\,{\sc iv} emission ratio are discussed in the context of nova
explosions, contamination of the secondary star and accretion of nova
abundance-enriched matter back to the white dwarf via the accretion disk
or as a descendant of a precursor binary that survived thermal timescale
mass transfer. The scenario involving pollution of the secondary by past
novae may be supported by the possible presence of a nova
remnant-like nebula around EY Cyg.

\end{abstract}

Subject Headings: Stars: cataclysmic variables, white dwarfs, Physical
Processes: accretion, accretion disks

\section{Introduction}

Dwarf novae are a subset of cataclysmic variables consisting of a white
dwarf accreting matter from a Roche lobe-filling, low-mass secondary star.
They undergo outbursts (brightness increases of 3-7 magnitudes) lasting
for days to weeks separated by intervals of quiescence lasting weeks to
years. The outbursts are thought to be caused by the release of
gravitational potential energy when matter in the disk periodically
accretes onto the white dwarf. The trigger for the accretion episode is
widely held to be the disk instability mechanism (e.g., Cannizzo 1998 and
references therein). During outburst, these systems are dominated by a
luminous accretion disk, a high rate of accretion, and wind outflow. The
spectra during quiescence are more difficult to interpret and generally
exhibit a mix of absorption and emission features with a continuum energy
distribution which may or may not be consistent with an accretion disk.

In this paper, we utilize composite white dwarf photosphere plus accretion
disk synthetic spectra to model the poorly understood far UV spectra of
dwarf novae in quiescence. Our approach is described in Winter \& Sion
(2003) and was also applied to the HST spectra of RX And (Sion et al.
2001; Sepinsky et al. 2002). Our approach is most useful for dwarf novae
in which we cannot easily identify the dominant light source during
quiescence in the far UV.  For the vast majority of dwarf novae, we do not
know the accretion rate during quiescence, nor is the spectrum of the
underlying white dwarf easily identifiable. Among the key questions we
seek to answer are: What is the accretion rate during quiescence? Is this
rate of accretion consistent with the disk instability model? How much
flux does the white dwarf contribute to the FUV?  How much flux does the
accretion disk contribute to the FUV? How hot is the white dwarf?

With these goals in mind, we expand our investigation of systems in
quiescence, with the selection of the U Gem-type dwarf novae, EY Cygni. It
exhibits a composition puzzle shared by only a few other non-magnetic and
magnetic CVs. It is well-known that the most prominent and common emission
line in the far UV spectra of dwarf novae is almost always C\,{\sc iv}
(1548 \AA, 1550 \AA). EY Cyg appears to be an exception to this rule
because it has very strong N\,{\sc v} emission and weak or absent C\,{\sc
iv}, as first reported by Winter \& Sion (2001). Further evidence of this
was seen in a recent study conducted by G\"{a}nsicke et al. (2003), where
examinations of HST/STIS spectroscopic snapshots of both BZ UMa and EY Cyg
revealed both systems to have among the largest N\,{\sc v}/C\,{\sc iv}
ratios observed in CV systems. Could the depleted carbon be pointing to a
different thermonuclear history for these dwarf novae, related either to
donor star contamination by nova ejecta from the accreting white dwarf
(see for example Sion et al.  1998), or nucleosynthetic processing in a
donor secondary originally more massive than the accreting white dwarf
that underwent a short phase of thermal timescale mass transfer (Schenker
et al.2002)? Is there any property of these two systems which points to
the reason for the peculiar N\,{\sc v}/C\,{\sc iv} emission line flux
ratio?

The orbital period and recurrence time between outbursts of EY Cygni has
recently been re-determined (Echeverria et al. 2002; Costero et al. 1998).  
The new orbital period is 0.45932 days (compared with the previous value
0.2185 days reported by Sarna et al. 1995).  The recurrence time of its
outbursts, based upon long term AAVSO archival light curve data, is 2000
days. Its light curve is highly variable and very unusual in that it
appears to be shifted by 0.5 phase relative to usual dwarf nova light
curves with similar parameters. All of this prompted Tovmassian et al.
(2002) to obtain direct images of EY Cyg in H$\alpha$ to search for a
nebulosity or common envelope.  From the radial velocity study (Echeverria
et al. 2002) and the light curve estimate (Tovmassian et al. 2002, Costero
et al. 1998, Smith et al. 1997), the inclination angle was found to be
very low.  They constrain $i$ to be in the range of 13 to 17 degrees.

An integral part of our synthetic spectral analysis is to derive a
distance fom the model fitting and compare it to estimated or derived
distances to these systems as a consistency constraint on the goodness of
fit.  Distances can be derived from the absolute magnitude at outburst
(M$_{v(max)}$) versus orbital period relation of Warner (1995). This
yields 505 pc for EY Cyg. Recent FGS parallaxes of dwarf novae by Johnson
et al. (2003) have yielded a new relation between the M$_{v(max)}$ and
orbital period, viz., M$_{v(max)}$ = 5.92 - 0.383 $P_{\rm orb}$ (hrs).
However, for EY Cyg, this relation yields 871 pc, which may be spurious
due to EY Cyg's period being far above the upper boundary of the period
gap.  A reasonable upper limit (700 pc) to the distance of EY Cyg is
imposed by the fact that it is not embedded in the Cygnus Superbubble (see
below) while a reasonable lower limit (250-300 pc) is imposed from the
spectral type of the secondary star. For EY Cyg, Tovmassian et al. (2002)
find the distance to lie between 400 and 500 parsecs. We adopt the mean of
this range (450 pc) as the distance to EY Cyg.

The observed characteristics of EY Cygni are given in Table 1 where, by
column, we list:  (1) orbital period in days, (2) recurrence time between
outbursts in days (3) orbital inclination, (4) spectral type of the
secondary, (5) mass of the primary in solar masses (6) mass of the
secondary in solar masses, (7) apparent magnitude in outburst, and (8)
apparent magnitude in quiescence. The references to the entries are listed
below the table.\\

\begin{deluxetable}{lcrcccccll}
\tablewidth{0pt}
\tablecaption{EY Cygni Parameters
\label{tab:para}}
\tablehead{\colhead{$P_{orb}$}
&\colhead{$t_{rec}$}&\colhead{$i$}&\colhead{sp. type($M_{2}$)}&\colhead{$M_{1}$}&\colhead{$M_{2}$}&\colhead{$V_{max}$}&\colhead{$V_{min}$}}
\startdata                       
0.45932 & 2000 & 13-17  &  K7V & 1.26       & 0.59  &11.4&15.5 \\
\\
 \tablecomments{\small References:\\
EY Cyg: Sarna et al. (1995), Smith et al. (1997), Tovmassian et al. (2002),
Costero et al. (2002)\\
Magnitudes for EY Cygni taken from Ritter \& Kolb (2003).\\}
\enddata
\end{deluxetable}

\section{FUSE observation of EY Cygni in Quiescence}

We obtained a FUSE spectrum of EY Cygni on 16 July, 2003 during its
quiescence. The exposure time was 18,514 seconds through the LWRS
aperture. The weakness of C III (1175) is consistent with the C depletion
evident in the STIS spectrum (see below) and the IUE archival spectrum.

We also obtained an HST spectrum under the Cycle 11 snapshot program led
by one of us (BTG). The details of the spectrum have already been
published and can be found in G\"ansicke et al. (2003).

\section{Ground-based direct imaging}

The ground-based observations of EY Cyg were done at the 3.5\,m
WIYN\footnote{The WIYN Observatory is a joint facility of the University
of Wisconsin-Madison, Indiana University, Yale University, and the
National Optical Astronomy Observatory.} telescope at Kitt Peak. The
imaging camera with 2K$\times$2K CCD and a set of nebular filters was
used. The observations in the H$_{\alpha}$ filter and corresponding narrow
band continuum were obtained on September 12, 2001. The other set of
observations, in filters S\,{\sc ii} and O\,{\sc iii}, were obtained one
year later on September 3, 2002 on the same telescope with the same
instrument. Images in each filter were obtained in multiple short
exposures in order to avoid saturation by bright stars. This method of
combining multiple images also helped to get rid of cosmic rays in the
best manner. The seeing was around 0.6 arcsec on average during both sets
of observations. The seeing in continuum images was slightly worse than in
H$\alpha$. In order to subtract continuum images from the emission line,
we had to do a Gaussian filtering to degrade the H$_{\alpha}$ images to the
level of continuum (about 0.1 arcsec). For the S\,{\sc ii} images we used
the scaled background image of the previous year, since no continuum image
was available in 2002. The table of optical observations is presented in
Table \ref{tab:optobs}.  The data reduction of mosaic images, including
bias processing, flatfielding, combining multiple exposures, and continuum
subtraction, was done using the MIDAS astronomical package.

\begin{deluxetable}{lcccc}
\tablewidth{0pt}
\tablecaption{Log of the Optical Observations of EY Cyg\label{tab:optobs}}
\tablehead{\colhead{UT Date}&\colhead{Band}&\colhead{Filter \#}
&\colhead{$t_{exp}$  sec.}&{Number of exp.}}
\startdata                                       
Sept 12 2001  & H$\alpha$  & W015    & 300  & 5\\
Sept 12 2001  & Cont.      & KP1494  & 300  & 2\\
Sept 03 2002  & S\,{\sc ii}       & W017    & 600  & 3\\
Sept 03 2002  & O\,{\sc iii}      & KP1467  & 600  & 2\\
\enddata
\end{deluxetable}

The surroundings of EY Cyg show complex H$\alpha$ clouds common to Cygnus.  
The images of EY Cyg in the H$\alpha$ filter are presented in Fig 1. Fig
1a presents a larger area around the object without continuum
substraction. This panel also shows the immediate surroundings of EY Cyg
after continuum substraction.  The star is situated at the rim of a very
circular ring of H$\alpha$ emission. We checked the possibility of EY Cyg
having a proper motion that could move it to the center of that ring in a
reasonable time stretch, but the position of the star does not change
within the limits of the measurements in the past 50 years. It is very
unlikely then that this ring is associated with our object.  The other
suspicious structure detected is in the form of a faint arc in the
immediate surroundings of EY Cyg, shown in Fig 1b. It is not homogeneous.
The size is about 25 arcsec. The lower limit of the distance to EY Cyg
from the secondary spectral class is at least 250 pc.  The upper limit is
probably 700 pc, otherwise the object would be embedded in the background
Cygnus superbubble and could hardly sustain its X-ray luminosity
(Bochkarev \& Sitnik 1985). Therefore, a reasonable distance estimate to
EY Cyg would be around 400-500 pc; this is also supported by the spectral
class of the secondary. In this case, the arc would be comparable in size
to the known nova shells of DQ Her and/or GK Per.  The morphology of this
formation is a bit different from larger scale strips crossing the region.  
Moreover, EY Cyg is situated in the center of the arc, suggesting that it
may be associated with the object. The images in S\,{\sc ii} show similar
but much fainter structure around the star.  We were not able to detect
any emission in O\,{\sc iii} from EY Cyg or from the larger area. The arc
in S\,{\sc ii} is only marginally seen, but its form coincides with that
of the arc in H$\alpha$, which confirms its reality and suggests a
possible association with EY Cyg. However, we cannot as yet 
definitively prove the association of the arc with the star.

\section{Synthetic Spectral Fitting}

Model spectra with solar abundances were created for high gravity stellar
atmospheres using TLUSTY (Hubeny 1988) and SYNSPEC (Hubeny \& Lanz 1995).
We adopted model accretion disks from the optically thick disk model grid
of Wade \& Hubeny (1998).  Using IUEFIT, a $\chi^{2}$ minimization
routine, both $\chi^{2}$ values and a scale factor S were computed for
each model.  The scale factor, normalized to a kiloparsec, can be related
to the white dwarf radius through: $F_{\lambda(obs)} = 4 \pi
(R^{2}/d^{2})H_{\lambda(model)}$, where d is the distance to the source, H
is the Eddington (surface) flux and $S = 4 \pi (R^{2}/d^{2})$. For each
system, the best-fitting accretion disk model was combined with the
best-fitting photosphere model to determine the contribution of both the
accretion disk and the white dwarf. For this combined white dwarf plus
disk model fitting, we took the best-fitting white dwarf model and
combined it with the best-fitting disk model, but, in effect, modulated
the accretion rate of the best-fitting disk model by multiplying the disk
fluxes by a small numerical factor in the range 0.1 to 10.0. In this
relatively narrow range, the scaling of the disk fluxes is roughly linear
so the scaling approximates varying the accretion rate by a small
increment and examining how it affects the $\chi^{2}$ value. We then found
the best-fitting composite white dwarf plus disk model based upon the
minimum $\chi^{2}$ value achieved and consistency of the scale
factor-derived distance with the adopted distance for EY Cyg.

The overlap in wavelength between the FUSE and STIS spectrum is 1150A to
1182A. The flux levels between the two spectra are mismatched by a factor
of 1.85 with the FUSE spectrum having a higher flux level than the STIS
spectrum. In order to get the best match of the spectra in the overlap
region, we raised the entire STIS spectrum by a factor of 1.85, but
eliminated the overlap region in the STIS spectrum and instead used the
FUSE overlap fluxes in that range since they are more reliable there.
While we cannot rule out that the mismatch is due to a STIS calibration
problem, we believe it is more likely that EY Cyg was in a lower accretion
state when the STIS spectrum was taken. Unfortunately, we have no optical
observations which document brightness differences between the times of
the STIS and FUSE spectra.  If we had divided the FUSE fluxes by 1.85 to
match the lower STIS fluxes, then a lower temperature from the overall fit
would have resulted. Hence, the temperature we have derived from raising
all of the STIS fluxes to match the FUSE fluxes in the overlap range can
be regarded as an upper limit.

For EY Cygni, we constructed a grid of white dwarf models with fixed
surface gravity log $g = 9$, T$_{eff}/1000 = 18,19, 20,....28$, Si and C
abundance 0.1, 0.2, 0.5, 1.0, 2.0 5.0 and V$_{rot} = 100, 200, 300, 400,
600, 800$. For the all of the accretion disk fitting, we adopted M$_{wd} =
1.2 M_\sun$ and $i = 18$ degrees in order to be consistent with the
published values and with the white dwarf gravity in the photosphere fits.
For the accretion belt fitting, taking $\sin 18 \sim 0.31, V_{belt} \sin i
= 2000$ km/s. This high velocity region is rotating near the Keplerian
velocity, much faster than the white dwarf rotation velocity of 400 km/s.
However, the high velocity region covers a relatively smaller fraction of
the white dwarf surface area and may, in reality, lie above the surface
(e.g. a hot rotating ring) rather than being right at the photosphere. For
the belt, we took log $g = 6$ and solar abundances for all elements. We
tried a range of accretion belt temperatures T$_{belt}/1000 = 30, 31, 32,
....50$ in steps of 1.

The results of our fitting of a single temperature white dwarf to EY Cyg
are somwhat encouraging in that the best-fitting model (i.e., least
$\chi^{2}= 1.233$) yielded a lower distance (301 pc) than our adopted
distance of 450 pc but not unreasonably lower. We find that the if the
FUSE +STIS spectrum is due to the white dwarf alone, then T$_{eff} =
24,000$K, log $g = 9.0$, V sin i = 600 km/s with Si abundance = 0.1 x
solar and C = 0.2 x solar. Even though the fitting yielded a rotational
velocity, this value must be viewed with caution due to the lack of
well-defined line profiles in the spectra. Since the FUSE and STIS spectra
were mismatched by a factor of 1.85 in the overlap region, the temperature
we have derived should be considered an upper limit (see above). This
best-fitting model is displayed in figure 2. Table 3 gives the fitting
parameters, where, by column, we list (1) is the model type, (2) is the
effective temperature, (3) the stellar surface gravity, (4) the rotational
velocity in km/s, (5) silicon abundance, Si/H, relative to solar, (6)
carbon abundance, C/H, relative to solar (7) the reduced $\chi^{2}$ value,
(8) the scale factor, (9) the distance in parsecs.
  
\begin{center}
\begin{tabular}{lcccccccc}
\multicolumn{9}{c}{Table 3:
EY Cygni Photosphere Fitting Results}\\ \hline\hline
Model &  T$_{eff}$ &  log g & Vsini&  Si&   C &   $\chi^{2}$&  S & d\\ \hline
WD &  24,000K&  9.0 &  600 &  0.1&  0.2 &  2.133&   $3.9\times 10^{-4}$& 302\\\hline
\end{tabular}
\end{center} 
        
On the other hand, we tried accretion disk model fits alone. The best
fitting disk model with M$_{wd} = 1.2 M_{\sun}$ and $i = 18$ degrees was for an
accretion a rate of $10^{-10} M_{\sun}$/yr had a $\chi^{2} =2.148$, almost
indistinguishable from the the white dwarf fit. However, the scale factor
of the disk fit yielded a distance = 1.8 kpc! 

The disk model plus white dwarf giving the minimum $\chi^{2}$ value was exactly
the accretion rate derived from the lone accretion disk fit. In the
combined fit, the accretion disk accounted for 96\% of the FUV flux while
the white dwarf accounted for 8\% of the FUV flux. However, the distance
derived from the scale factor of the combined fit was 1,060 pc,
substantially larger than than our adopted most probable distance of 450
pc. 

Finally, we attempted fits involving two-temperature white dwarfs, in
which a hotter, rapidly spinning region of small surface area (e.g., an
accretion belt) is combined with a white dwarf rotating more slowly with a
lower temperature. We also constructed accretion belt models with log $g =
6$, V$_{belt} = 3400$ km/s, T$_{belt}/1000 = 30,31, ...50$.  The
best-fitting white dwarf plus belt model yielded the lowest $\chi^{2}$
value (2.03) of all of the above single and two-component fitting
experiments and a distance = 351 pc, within a 100 pc of the adopted
distance of 450 pc. The white dwarf contribution to FUV flux is 37\% while
the accretion belt contributes 63\% of the FUV flux. The accretion belt
covers 27\% of the white dwarf surface area, has a belt temperature of
36,000K, log $g = 6$ and V$_{belt}$ sin $i = 2000$ km/s. The best-fitting
accretion belt plus white dwarf model is displayed in figure 3 and the
fitting parameters are given in Table 4 where we list by column: (1) the
model type (2) the white dwarf mass (3) the WD T$_{eff}$ (4) the accretion
belt temperature (5) accretion belt velocity (6) percentage flux
contribution of the WD (7) percentage flux contribution of the belt (8)
the reduced $\chi^{?}$ value (9) scale factor and (10) the distance in
parsecs.

\begin{center}
\begin{tabular}{lccccccccc}
\multicolumn{10}{c}{Table 4:
EY Cygni White Dwarf plus Accretion Belt Fitting Results}\\\hline\hline
Model&  M$_{wd}$ & T$_{eff}$ & T$_{belt}$ & V$_{belt}$& Flux (WD)& Flux (Belt)& $\chi^{2}$& S& d\\  \hline
WD+Belt& 1.2 &  22,000K& 36,000K& 2000 & 37\% &    63\% &      2.033 &
$2.88\times 10^{-4}$& 351 \\ \hline
\end{tabular}
\end{center}

\section{Conclusions}

Based upon the fitting experiments described in the preceding section, it
seems clear that the FUSE plus STIS spectrum of EY Cygni in quiescence is
best-represented by either a single temperature white dwarf with T$_{eff}
= 24,000$K or by a white dwarf with T$_{eff} = 22,000$K and an accretion
belt having T$_{belt} = 36,000$K and V$_{belt}$ sin $i = 2000$ km/s. Since
the latter fit has a lower $\chi^{2}$ and yields a distance closer to
450pc, it is favored over the single temperature white dwarf fit. The
accretion belt is the dominant source of FUV flux from the FUSE range
through the STIS range with the cooler white dwarf latitudes contributing
less than 40\% of the FUV flux.

In EY Cygni, which lies above the period gap, the $T_{\rm eff}$ value
appears to be lower than the typical 29,000 to 35,000K of other U Gem and
Z Cam systems above the gap. Could there be any connection between the
high mass white dwarf, lower T$_{eff}$ of EY Cygni and the abundance anomaly
seen in the far-UV spectra? With a higher white dwarf mass and lower
$T_{\rm eff}$ value, it is possible that the abundance anomaly (excess N,
depleted C) may be associated with a greater system age (longer cooling
time of the white dwarf) than other dwarf novae. If indeed the white dwarf
core has cooled for a longer timescale, then the equilibrium $T_{\rm eff}$
value in response to accretion is expected to be lower. This might suggest
that an older system age and massive white dwarf allows for more classical
novae episodes, hence more CNO nucleosynthetic processing, consequently
more contamination of the secondary's atmosphere during the brief common
envelope stage of each nova. Such systems may be more likely to exhibit
the apparent excess N and depleted C associated with CNO processing. Our
direct images with a narrow H$\alpha$ filter of EY Cyg and its
surroundings show the nova-like appearing envelope around it, but were
inconclusive as to whether the nebular emission around the object is
associated with it.

An alternative, more promising picture has emerged however. Recent work by
G\"ansicke et al. (2003) suggest that the N/C anomaly seen in the dwarf
novae BZ UMa, EY Cyg, 1RXS J232953.9+062814, and now CH UMa (see Dulude \&
Sion 2002), may have its origin in a CV with an originally more massive
donor star ($M_{2} > 1.5 M_{\sun}$) which survived thermal time scale mass
transfer (Schenker et al. 2002 and references therein). In such a system,
the white dwarf would be accreting from the peeled away CNO-processed core
stripped of its outer layers during the thermal timescale mass transfer.

\section{Acknowledgments}
Summer undergraduate research support for L.W. and J.U. from the 
Delaware Space Grant Consortium is gratefully acknowledged.
This work was also supported by NSF grant AST99-09155 and by NASA grant
NAG5-11578 to Villanova University.
G.T. and M.O. acknowledge a grant from the University of Wisconsin Graduate
School, and financial support from the Ministry of Instruction, Research
and Universities (MIUR) of Italy "COFIN" program.
BTG was supported by a PPARC Advanced Fellowship,

\clearpage

Fig.1a- A larger area than on the right is presented 
without continuum subtraction, showing the surroundings of EY Cyg.  \textit{CV position represented by \emph{X}.}

Fig.1b- A blow-up of the continuum subtracted image of EY Cyg. A faint 
arc crossing the larger scale strips of nebular emission
can be seen to the north-east of the star.

Fig.2- The best-fit single temperature white dwarf model to the 
combined FUSE + HST/STIS spectrum of EY Cyg during quiescence. The white
dwarf model has $T_{\rm eff} = 24,000$K, log $g = 8.96$, rotational velocity
Vsini = 600 km/s, with Si abundance = 0.1 x solar and C = 0.2 x solar. 

Fig.3- The best-fit combination of white dwarf plus accretion belt
synthetic fluxes to the spectrum of EY Cyg during quiescence. The white
dwarf model has $T_{\rm eff} = 22,000$K, log $g = 8.9$ and the accretion
belt has log $g = 6$, Vsini = 2000 km/s and covers 27\% of the WD surface
area. The top solid curve is the best-fitting combination, the dotted
curve is the white dwarf spectrum alone and the dashed curve is the
accretion belt synthetic spectrum alone. In this fit, the accretion belt
contributes 63\% of the far UV flux, and the white dwarf contributes 37\%
of the flux.

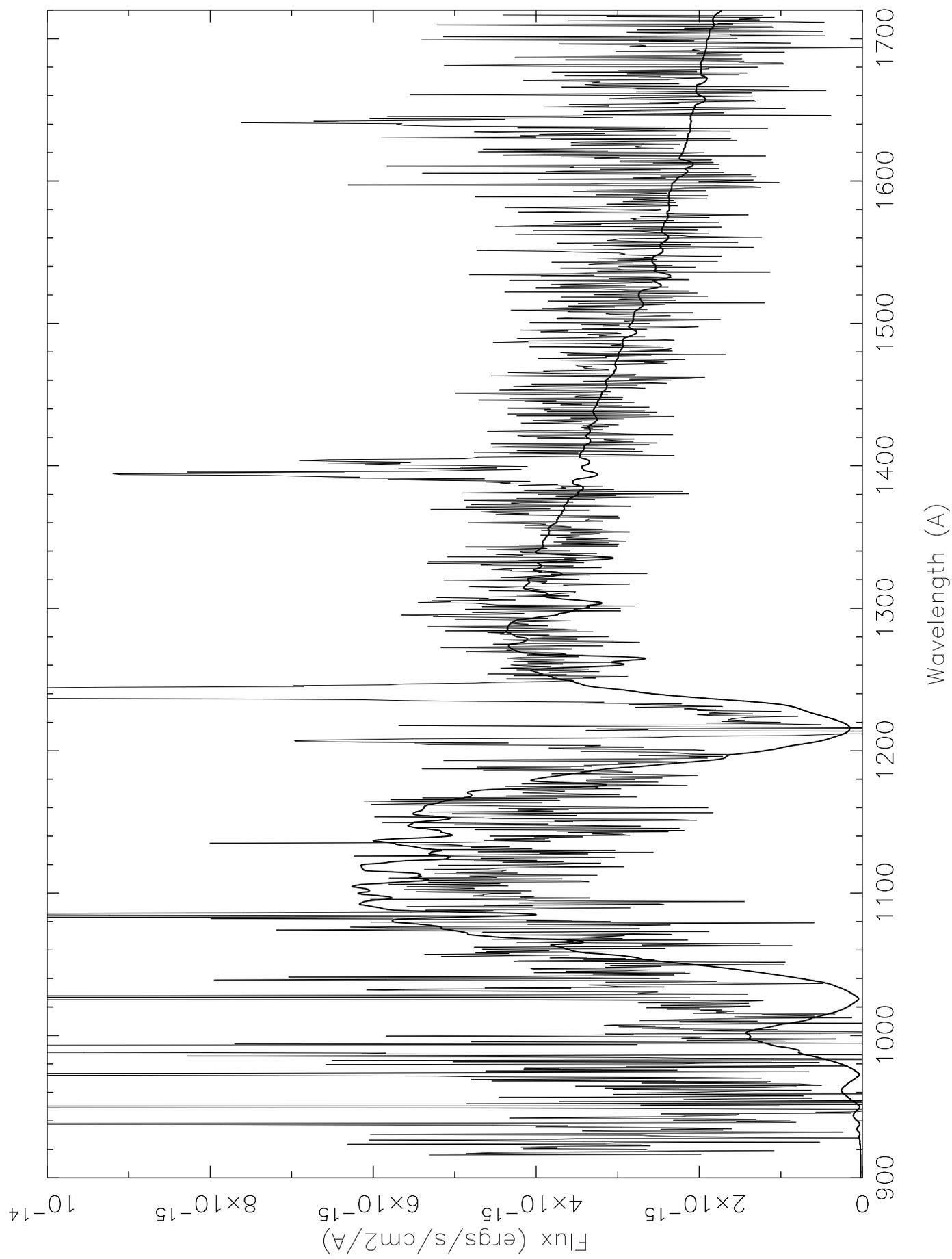
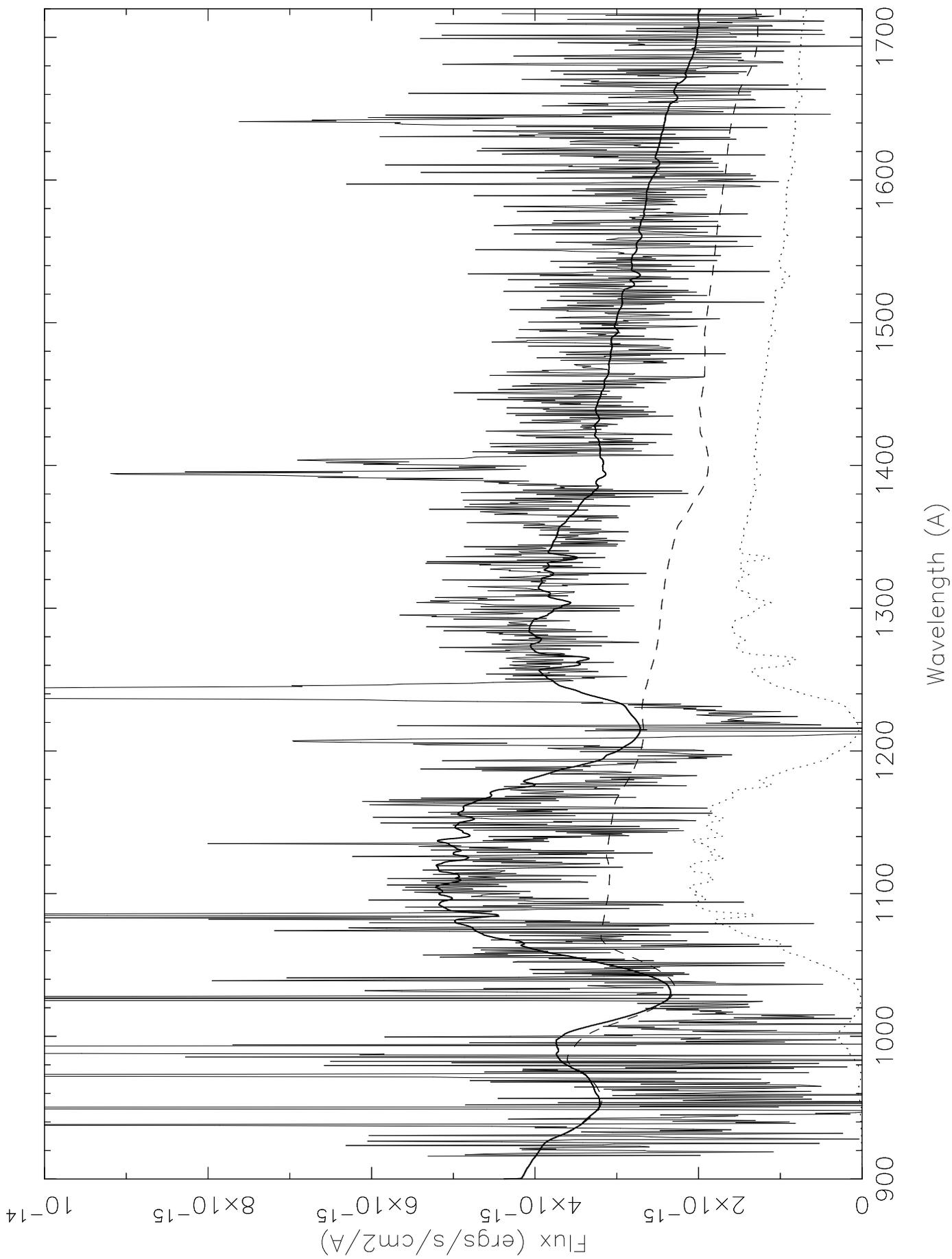
\end{document}